# Modeling the Sea-Level Change from U.S. Vehicle Emissions

Tony Wong
School of Mathematics and Statistics
Rochester Institute of Technology
tony.wong@rit.edu

## Abstract

Recent U.S. Environmental Protection Agency (EPA) analyses have argued that greenhouse gas emissions from U.S. on-road vehicles contribute negligibly to global mean sea-level rise (GMSLR). Here, I replicate and extend the EPA's modeling framework using the FaIR climate model coupled with the BRICK sea-level model, incorporating a probabilistic weighting approach and a longer model timescale to better represent joint climate-sea-level uncertainty. In addition to the baseline SSP2-4.5 scenario and an EPA-consistent emissions reduction case, I examine alternative scenarios reflecting stalled technological progress and a counterfactual pre-regulation vehicle fleet. Results reproduce EPA estimates of approximately 1-2 cm of GMSLR reduction by 2100 under vehicle emissions mitigation but show that these differences grow substantially over multi-century timescales, exceeding 6 cm by 2200. Downscaling to U.S. coastlines reveals larger local effects, particularly along the Gulf of Mexico Coast. These findings highlight the long-term and regionally amplified benefits of emissions reductions from the transportation sector.

## 1 Introduction

In February 2026, the United States Environmental Protection Agency (EPA) repealed its 2009 Greenhouse Gas Endangerment Finding and along with it, its Motor Vehicle Greenhouse Gas Emissions Standards Under the Clean Air Act.[1] The justification for this relied partially on computational modeling to show that the consequent sea-level rise associated with greenhouse gas (GHG) emissions from U.S. on-road vehicles is *de minimis* - that is, too small to matter. Included in the final rule is a "Technical Memo on: Temperature, CO2 Concentration, and Sea Level Rise Impacts of Greenhouse Gas Emissions from U.S. Motor Vehicles for the "Rescission of the Greenhouse Gas Endangerment Finding and Motor Vehicle Greenhouse Gas Emission Standards Under the Clean Air Act" Final Rule[2] (henceforth referred to as "EPA Technical Memo"). The EPA Technical Memo provides details on the modeling done, including data tables giving their results. EPA's estimated reduction in global mean sea-level rise (GMSLR), assuming emissions associated with all U.S. on-road vehicles are removed beginning in 2027, is 1.4 cm over the 2027-2100 time period, with a 95% credible interval of 0.39-4.77 cm. The EPA modeling chain used the Building Blocks for Relevant Ice and Climate Knowledge (BRICK) sea-level model[3], of which I am one of the two original lead authors. I continue to manage the development of BRICK.

Here, I examine the baseline SSP2-4.5 GHG emissions scenario and several reduced emissions scenarios mirroring the cases in the EPA Technical Memo. I connect the temperature and ocean heat changes resulting from these scenarios to global mean sea-level rise, and subsequently to local mean sea-level rise (LMSLR) for the United States coastline. These projections can facilitate more targeted evaluations of the economic impacts to the United States associated with reduced GHG emissions from U.S. on-road vehicles. Further, there are a number of modeling choices in the EPA Technical Memo that I seek to clarify and make transparent in the work presented here.
- First: GMSLR and GMST were used instead of local hazards. As these are global means, they may differ from the actual hazards faced by the United States' coasts. Here, I downscale the GMSLR from BRICK to LMSLR for the U.S. coasts. I also aggregate over the U.S. Gulf of Mexico Coast to capture a region more geographically homogeneous.
- Second: It is not clear which version of the BRICK model was used. Given the timing of the EPA Final Rule and accompanying Technical Memo (February 2026), and an interest in reproducing the EPA's general methods for purposes of understanding the Agency's analysis, I employ a model version of BRICK that was released in February 2025; the next model version was released in December 2025.
- Third: It is not clear which version of calibrated BRICK model parameters were used. Again, given the timing of the EPA modeling work, I employ a set of BRICK model parameters that were released alongside the model version noted above. Other structural choices for the model version and calibrated parameters are possible, and reasonable. The choices I made here are in an effort to reproduce the EPA modeling chain as accurately as possible based on the details provided in the EPA Technical Memo. I note that the Technical Memo does not include any supplemental files such as the computer codes used to perform the analysis. This standard scientific practice of providing underlying data and code builds and maintains trust in the scientific enterprise.
- Fourth: The EPA Technical Memo treats the combined FaIR-BRICK simulations as equally likely (p. 4). The models were calibrated separately, then combined, so the simulations constitute samples from a probability distribution that is distinct from the true FaIR-BRICK joint posterior distribution. Specifically, the FaIR parameters are sampled from their marginal posterior distribution, as are the BRICK parameters. I use a model weighting approach to reweight the FaIR-BRICK simulations according to how well they match observational data and account for the differences in sampling distributions.
- Fifth: Due to the multi-century scale of the sea-level response to GHG emissions and consequent global warming[4–7], a longer simulation period than through the year 2100 is warranted. Here, I use a simulation period through the year 2200 to more fully capture the sea-level response to reduced GHG emissions from U.S. on-road vehicles.

---

## 2 Modeling Workflow

### 2.1 Sea Level Modeling

I use the MImiBRICK[8] v1.2.0 implementation of BRICK, along with associated sets of calibrated parameters for BRICK[9]. MimiBRICK is an implementation of the BRICK sea-level model in Julia, in the Mimi integrated modeling framework[10]. MimiBRICK v1.2.0 is a model tag released in February 2025, while the following tag (v1.2.1) was released in December 2025. The MimiBRICK Zenodo repository contains several calibrated parameter sets, corresponding to multiple model configurations. The specific parameters file used in this work is "parameters_subsample_brick.csv", which was generated by calibrating BRICK in a "standalone" format, forced by a single trajectory of global mean surface temperature (GMST) and ocean heat uptake. Based on the modeling details provided in the EPA Technical Memo, this is my best guess at the BRICK configuration used by EPA.

In order to replicate the modeling workflow as closely as possible, based on the details provided in the EPA Technical Memo, I use input for GMST and ocean heat uptake from the Finite Amplitude Impulse Response climate model[11,12] (FaIR; version v2.2.0), using SSP2-4.5 radiative forcing and GHG emissions and a set of 841 calibrated parameters (v1.4.1)[13]. FaIR is an open-source, reduced-complexity climate model, designed for probabilistic projections of global temperature and atmospheric GHG concentrations from emissions.

In addition to the baseline SSP2-4.5 scenario, which mirrors scenario #1 in the EPA Technical Memo, three additional scenarios are produced. Scenario A here reconstructs scenario #2 in the EPA modeling and captures EPA's projected level of on-road vehicle emissions in the United States, absent GHG regulations. Emissions trajectories under scenario A interpolate the sparse data points provided in the EPA Technical Memo (2027, 2050, and 2100) and hold emissions constant after the year 2055. Scenario B here represents a case in which U.S. vehicle technology stays at model year 2025 levels (the most recent year with reported data) and rates of adoption of electric vehicles remain constant, but total vehicle emissions evolve through changes in vehicle miles traveled. Scenario C here represents a counterfactual scenario without modern vehicle GHG regulations, fixing emissions rates at circa 2009 levels with no electric vehicle adoption and allowing total emissions to grow with increasing vehicle use.

In this way, scenarios B and C represent "today's fleet" and a "pre-GHG protection fleet", respectively, while scenario A mirrors the EPA Technical Memo scenario #2. Echoing the EPA Technical Memo and Final Rule, I note that this approach models removing all emissions associated with U.S. on-road vehicles, such that the projections presented in the EPA Technical Memo and here constitute in some sense an upper bound on the reductions in sea-level rise as a result of any specific set of GHG standards. However, the assumption of SSP2-4.5 as the baseline scenario neglects the fact that this "middle of the road" or "maintain current policy" scenario relies on actually maintaining the decarbonization efforts that were underway a decade ago when the Shared Socioeconomic Pathways were developed[14,15]. These factors will have compensatory effects.

For each of the 841 emissions trajectories, the global mean surface temperature and ocean heat uptake model output from FaIR serves as input to BRICK. Output from BRICK includes global mean sea-level rise, and the contributions to GMSLR from the Greenland and Antarctic ice sheets, glaciers, thermal expansion, and land water storage. To facilitate reproducibility, I set random number seeds and save the indices of the BRICK ensemble members sampled from the larger dataset of 10,000 calibrated parameter sets. However, due to insufficient details provided to exactly replicate the experimental set-up from the EPA Technical Memo, it is not expected that emissions trajectories or the resulting projections of global temperatures or sea levels will precisely match those presented in the Technical Memo.

## 2.2 Model Calibration

The BRICK model parameters in the dataset noted above were calibrated based on the model-data match to observational datasets for the major components of global mean sea-level change: the Greenland and Antarctic ice sheets, glaciers, thermal expansion, and land water storage. The Bayesian model calibration algorithm used is described in detail elsewhere[3,16,17], and the resulting distributions for sea-level projections are consistent with the Intergovernmental Panel on Climate Change's Sixth Assessment Report (IPCC AR6)[18]. Importantly, BRICK contains a simple module to account for potential "low confidence" but high-impact Antarctic ice sheet processes that can contribute substantially to sea-level change in the coming centuries[16]. Thus, projections using BRICK are expected to fall on the higher side of probable ranges for sea-level rise for the latter half of the 21st century and beyond, particularly for higher GHG emissions scenarios. In SSP2-4.5, these fast dynamical contributions to sea level from the Antarctic ice sheet can occur by the year 2100[16].

The parameter datasets that EPA used for FaIR and BRICK were calibrated independently of one another, then combined to form paired (concomitant) parameter sets for a coupled FaIR-BRICK model, wherein temperatures and ocean heat uptake output from FaIR serves as input to BRICK. The EPA Technical Memo asserts that all parameter sets and resulting simulations are equally likely. From a model calibration and statistical modeling standpoint, this is not the case because the BRICK simulations were calibrated using different temperature and ocean heat input than that from FaIR. On the other hand, the simulations in the ensemble constructed here are intended to be samples produced from the joint distribution of FaIR-BRICK parameters.

As an example, a set of FaIR parameters that yields a warm simulation for temperature could be paired with a set of BRICK parameters that yields a low simulation for sea-level rise. This should be treated as having relatively lower probability than more compatible sets of FaIR and BRICK parameters. To account for the variation in the goodness-of-fit of the combined FaIR-BRICK parameters, I compute weights for each concomitant parameter set and resulting simulation and use these weights to compute weighted percentiles for resulting ensemble statistics (e.g., for global mean sea-level change).

$$log(w_i) = c \cdot (l(\theta_{i,FB}) - l(\theta_{i,B})) \qquad (1)$$

In Equation 1, $\theta_{i,B}$ refers to the i-th set of BRICK model parameters and $\theta_{i,FB}$ refers to the i-th set of FaIR-BRICK parameters (so the BRICK parameters are the same for both). In turn, $l(\theta_{i,B})$ refers to the value of the log-likelihood function when the BRICK model is run using its original temperature and ocean heat forcing and the i-th set of BRICK model parameters, and $l(\theta_{i,FB})$ refers to the log-likelihood value when the combined FaIR-BRICK model is run using the combined i-th set of FaIR-BRICK parameters. In Equation 1, c is a constant that is tuned to balance the influence of the best-fitting simulations against sampling from the full breadth of the approximate joint distribution of FaIR-BRICK simulations. The computed weights use a value for this "annealing" constant of c=0.000128. This yields an effective sample size of about 420, or roughly 50% of the original sample size. This is in line with guidelines and typical practice for importance sampling methods[19,20], where samples from one distribution are desired (e.g., the distribution of FaIR-BRICK parameters) but inaccessible, so they are approximated using samples from an easier-to-sample distribution (e.g., the marginal distributions of FaIR and BRICK parameters). Only the sea-level portion of the likelihood function is used because all 841 sets of FaIR parameters are known *a priori* to be well-calibrated to climate data; the BRICK parameters were calibrated to sea-level data, but the quality of the simulation is affected by changing the underlying temperature and ocean heat forcing data.

The log-weights from Equation 1 are then shifted by subtracting the maximum weight (to center them closer to 0 and avoid undue influence from the best-fitting simulations) and exponentiated to compute the unnormalized simulation weights, as shown in Equation 2.

$$W_i = exp[log(w_i) - max(log(w_i))] \qquad (2)$$

Normalized weights are obtained by dividing by the sum, as shown in Equation 3.

$$W_{i,norm} = W_i / \sum_{i=1}^{841} W_i \qquad (3)$$

**Figure 1:** Computed weights for each FaIR-BRICK simulation, based on the BRICK log-likelihood function. The dashed red line corresponds to equal weights of 1/841 ≈ 0.0012.

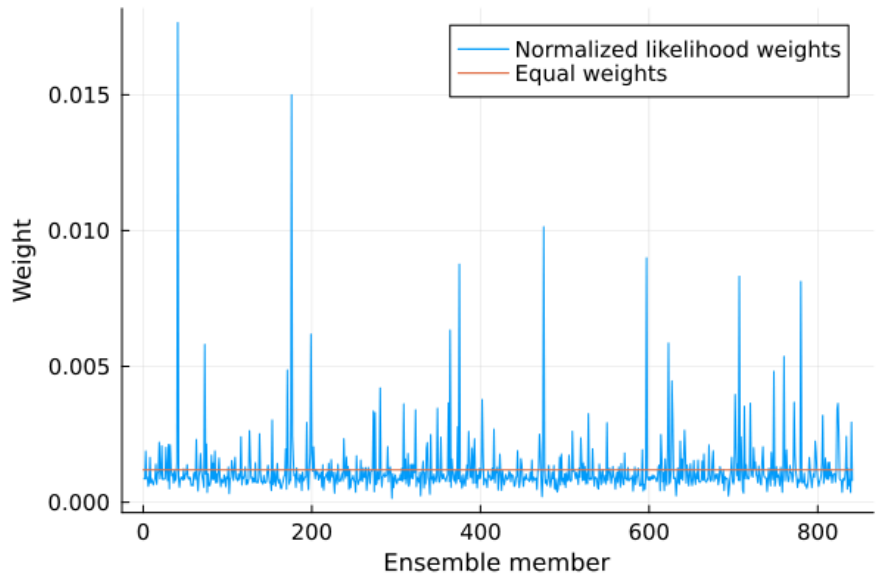

## 2.3 Local Hazards

The EPA Final Rule and Technical Memo use changes in GMST and GMSL as the measure of the impact of rescinding the vehicle emissions standards in the Clean Air Act. I use the BRICK sea-level output to estimate the associated local sea-level changes for the United States, which is better connected to the actual consequences to the U.S. of the policy change. In the case of sea-level rise, this is because local mean sea-level change can differ from global mean sea-level change due to differences in vertical land motion, ocean dynamics, and gravity effects from redistributing large amounts of ice/water[21].

I select two model simulations to downscale to local mean sea level. The first model simulation is the one with the highest normalized weight (Equation 3). I refer to this simulation the "MLE" simulation, as a maximum likelihood estimator is an approximate interpretation of how it was selected. I caution however that other simulations among the FaIR-BRICK ensemble also provide high-quality fits to the BRICK sea level calibration data, based on the model weights (Figure 1). The second model simulation that I downscale to local sea-level rise is the one that yields the median GMSLR in the year 2100 under the baseline scenario. These two simulations provide a reasonable representation of likely future local sea level rise for the United States coastline.

To compute the local mean sea-level rise (LMSLR) in the four scenarios described above, for each of the two model simulations chosen, I downscale the components of global mean sea-level change in each case to their effects on local mean sea level on a 1-degree latitude-longitude grid, using a set of well-established sea level fingerprints[22]. I compute the LMSLR for each of the 1,077 United States coastal segments from a common coastal database[23], over the 2010-2200 time horizon. To examine a specific geographic region with more similar properties, I also aggregate the localized sea-level rise over the 178 coastal segments of the United States Gulf of Mexico Coast.

# 3 Results

Where parenthetical uncertainty ranges are reported in the results, they are 95% credible intervals.

### 3.1 Global mean sea level

Results for global mean sea-level change relative to pre-industrial mean (1850-1900) are broadly consistent with those reported in the EPA Technical Memo in the baseline case (Table 1). In 2050, I find GMSLR of about 27.2 (18.4-53.6) cm, as compared to 38.9 cm reported in the EPA Technical Memo (relative to 1850-1900 mean). By 2100, GMSLR in the baseline case yields about 91.3 (42.5-179.4) cm, as compared to 94.3 (59.9-157.9) cm by the EPA's modeling. As expected, scenarios A, B, and C all match the baseline scenario in 2009, and scenarios A and B match the baseline case in 2027.

**Table 1:** Global mean sea-level rise relative to pre-industrial mean (1850-1900), by scenario. Shown are the ensemble projected median and 95% credible interval (cm), weighted as described in Sec. 2.2.

|      | **Baseline**        | **Scenario A**      | **Scenario B**      | **Scenario C**      |
|------|---------------------|---------------------|---------------------|---------------------|
| 2009 | 10.1 (6.0-15.3)     | 10.1 (6.0-15.3)     | 10.1 (6.0-15.3)     | 10.1 (6.0-15.3)     |
| 2027 | 15.6 (10.6-22.5)    | 15.6 (10.6-22.5)    | 15.6 (10.6-22.5)    | 15.5 (10.5-22.5)    |
| 2050 | 27.2 (18.4-53.6)    | 27.0 (18.4-53.5)    | 27.0 (18.4-53.5)    | 26.8 (18.3-52.2)    |
| 2100 | 91.3 (42.5-179.4)   | 89.7 (42.0-176.5)   | 89.5 (41.9-176.3)   | 87.5 (41.4-174.2)   |
| 2150 | 177.9 (68.6-330.0)  | 174.7 (67.2-324.9)  | 174.3 (66.9-323.8)  | 170.5 (65.7-318.5)  |
| 2200 | 268.5 (95.6-475.1)  | 260.4 (90.5-467.6)  | 259.1 (89.8-464.9)  | 252.6 (87.5-456.1)  |

Results for global mean sea-level change relative to 2027 (the first year of assumed emissions reductions) also agree well in the baseline scenario and scenario A, relative to the EPA Technical Memo (Table 2). By 2050, I find 11.4 (6.0-32.7) cm of GMSLR in the baseline case, as compared to 12.4 (9.4-20.3) cm in the EPA Technical Memo. By 2100, this GMSLR reaches 76.1 (29.9-160.7) in this work, and 69.5 (35.2-132.7) cm in the EPA modeling. The reduction in GMSLR in scenario A relative to the baseline case is 0.07 (0.04-1.04) cm in 2050 and 1.45 (0.40-5.16) cm in 2100, compared to the reductions of 0.09 (0.06-1.06) cm in 2050 and 1.40 (0.39-4.77) cm in 2100. Over the 22nd century, these benefits in terms of reduced GMSLR increase to 3.61 (1.16-12.73) cm and 6.39 (2.35-21.18) cm in 2150 and 2200, respectively, in scenario A, and 4.28 (1.43-17.32) cm and 8.32 (3.12-29.34) cm in 2150 and 2200 in scenario B.

**Table 2:** Global mean sea-level rise relative to 2027, by scenario. Shown are the ensemble projected median and 95% credible interval (cm), weighted as described in Sec. 2.2.

|      | **2027 Baseline**   | **Scenario A**      | **Scenario B**      |
|------|---------------------|---------------------|---------------------|
| 2050 | 11.4 (6.0-32.7)     | 0.07 (0.04-1.04)    | 0.07 (0.04-1.03)    |
| 2100 | 76.1 (29.9-160.7)   | 1.45 (0.40-5.16)    | 1.61 (0.44-5.82)    |
| 2150 | 163.6 (55.4-309.3)  | 3.61 (1.16-12.73)   | 4.28 (1.43-17.32)   |
| 2200 | 253.8 (82.1-454.1)  | 6.39 (2.35-21.18)   | 8.32 (3.12-29.34)   |

In (counterfactual) scenario C, reductions in GMSLR relative to baseline would reach 0.04 (0.03-0.08) cm by 2027, 0.29 (0.15-2.23) cm by 2050, 3.57 (0.91-9.65) cm by 2100, and 14.4 (5.17-60.3) cm by 2200 (Table 3).

**Table 3:** Global mean sea-level rise relative to 2009, for the baseline scenario and scenario C. Shown are the ensemble projected median and 95% credible interval (cm), weighted as described in Sec. 2.2.

|      | **2009 Baseline**    | **Scenario C**      |
|------|----------------------|---------------------|
| 2027 | 5.5 (2.5-9.3)        | 0.04 (0.03-0.08)    |
| 2050 | 17.0 (9.6-41.0)      | 0.29 (0.15-2.23)    |
| 2100 | 81.0 (33.3-169.0)    | 3.57 (0.91-9.65)    |
| 2150 | 167.8 (59.0-317.4)   | 8.25 (2.58-30.40)   |
| 2200 | 260.5 (85.6-463.4)   | 14.37 (5.17-60.30)  |

### 3.2 Local mean sea level

I compute the local mean sea-level rise for all U.S. coastal segments within the DIVA database[23]. This dataset divides the global coastline into 12,148 segments with a median length of about 17 km. Since the coastlines of the United States show variation in major drivers of risk (e.g., east versus west coasts of the continental U.S., also, Alaska, Hawai'i, and various U.S. outlying territories), I specifically examine the mean local sea-level change along the U.S. Gulf of Mexico Coast ("Gulf Coast"). For ease of visualization and interpretation, I only downscale the FaIR-BRICK ensemble members yielding the maximum likelihood weight ("MLE") and the median GMSLR in the baseline case in the year 2100 ("Med2100"). These two simulations both are reasonable structural choices for a single "best-fitting" representative model simulation and well-represent a likely range of anticipated uncertainty around the ensemble centers. I note that GMSLR was downscaled to local sea-level rise on a 10-year time step to match the dataset used to account for non-climatic factors affecting local sea levels[24–26]. Consequently, results in Table 4 are shown relative to the year 2030 instead of 2027 as in the EPA Technical Memo and Table 2.

**Table 4:** Baseline GMSLR and LMSLR for the U.S. Gulf of Mexico Coast (cm), shown relative to 2030 for consistency with how downscaling was done, which used a 10-year timestep. Gulf Coast baseline is the mean local mean sea level for all 178 US Gulf of Mexico coastal segments, relative to 2030, in each of the two downscaled simulations, MLE and Med2100. Scenarios A, B, and C are given as the reduction in LMSLR relative to the Gulf Coast baseline scenario.

|      | **GMSLR relative to 2030** | | **Gulf Coast baseline** | | **Scenario A** | | **Scenario B** | | **Scenario C** | |
|------|------|---------|------|---------|------|---------|------|---------|------|---------|
|      | MLE  | Med2100 | MLE  | Med2100 | MLE  | Med2100 | MLE  | Med2100 | MLE  | Med2100 |
| 2050 | 11.2 | 8.76    | 12   | 12      | 0.1  | 0.1     | 0    | 0.1     | 0.2  | 0.2     |
| 2100 | 47.4 | 78.3    | 53   | 92      | 2.8  | 3.7     | 2.8  | 3.8     | 3.4  | 6.5     |
| 2150 | 106.2| 164.3   | 121  | 190     | 12.5 | 4.8     | 15.1 | 5.1     | 20.9 | 8.7     |
| 2200 | 181.7| 249.7   | 205  | 286     | 20.1 | 6.2     | 26.5 | 7.3     | 55.5 | 12.1    |

The tendency of the Gulf of Mexico to experience higher LMSLR than GMSLR is evidenced by both the MLE and Med2100 simulations (Table 4, first 4 columns). The MLE simulation has slightly higher LMSLR than GMSLR in 2050 relative to 2030 (12 cm locally versus 11.2 globally). This difference grows to over 5 cm by 2100 and more than 20 cm by 2200. The Med2100 simulation displays an even higher rate of local sea-level rise compared to GMSLR. In the Med2100 simulation, by 2050, the Gulf Coast experiences more than 3 cm higher sea-level rise than global mean. By 2100, this difference increases to more than 13 cm and by 2200, more than 30 cm higher LMSLR than GMSLR in the Med2100 simulation. While land subsidence is responsible for much of the comparatively higher LMSLR for the Gulf Coast, the relative contributions from the major ice sheets, Greenland and Antarctica, also play a key role. Specifically, due to gravitational effects from melting large amounts of ice, contributions to GMSLR from the Greenland ice sheet serve to lower local mean sea level along the Gulf Coast, whereas contributions from the Antarctic ice sheet raise local mean sea level there. Since the Med2100 simulation sees a much higher sea level contribution from the Antarctic ice sheet, LMSLR is also relatively higher compared to the MLE simulation.

Also owing to the relatively larger sea level contribution from the Antarctic ice sheet, the Med2100 simulation shows lower reductions in LMSLR in the three reduced emissions scenarios when compared to the MLE simulation (Table 4, right 6 columns). Reductions in LMSLR for the Gulf Coast in scenario A for the two simulations considered here span 2.8-3.7 cm in 2100 to 6.2-20.1 cm by 2200. This exceeds the median reduction in GMSLR of 1.4 cm by 2100 reported in the EPA Technical Memo, indicating that local benefits for the U.S. Gulf Coast can be substantially larger than those suggested by global mean values alone. Scenario B yields similar reductions in 2100 (2.8-3.8 cm of reduced LMSLR) but larger benefits by 2200 (7.3-26.5 cm). Scenario C shows even greater and earlier reductions in LMSLR, with 0.2 cm in 2050, 3.4-6.5 cm by 2100, and 12.1-55.5 cm by 2200.

## 4 Discussion

I have implemented a modeling workflow that mirrors the sea-level projections workflow presented in the EPA Technical Memo as closely as possible given the details provided, while also improving the modeling workflow by taking into account variation in the quality of ensemble members through a weighting approach. I find projections of GMSLR that are broadly consistent with the projections as presented in the EPA Technical Memo, particularly for GMSLR by 2100 in the baseline scenario and in the reduced emissions scenario A. By downscaling the GMSLR projections to LMSLR for the entire U.S. coastline and the U.S. Gulf of Mexico Coast, I connect these sea-level projections to local coastal hazards more specific to the United States. These LMSLR projections, particularly for scenario A relative to the baseline scenario, demonstrate that vehicle emissions reductions via the Clean Air Act disproportionately benefit the U.S. Gulf Coast (Table 4).

For global mean sea level, the difference between the baseline SSP2-4.5 scenario and scenario A (EPA's scenario #2) is modest in the near-term, but grows over time, reflecting the long timescale of the sea-level response to changing emissions and temperature. By 2100, removing U.S. on-road vehicle emissions yields a median reduction of about 1.45 cm of GMSLR, but this effect more than quadruples to over 6 cm by 2200 (Table 2). Scenarios B and C also produce greater GMSLR reductions by 2100, and by the year 2200, these benefits grow to over 8 cm in scenario B and over 14 cm in scenario C (Tables 2 and 3). These results highlight the importance of considering the long-term consequences of continued emissions, even at marginal GHG levels that may seem too low to matter.

In the downscaling to local coastal hazards, for ease of interpretation, I use just two model simulations among the 841-member ensemble to characterize uncertainty. These are the simulation with the maximum likelihood weight among the ensemble (MLE) and the simulation yielding the median GMSLR in the year 2100 in the baseline scenario (Med2100). Both simulations were chosen based on their match to central tendency in the ensemble. Consequently, they may well underestimate the true breadth of potential future LMSLR for the U.S. Gulf Coast, even though the uncertainty ranges in Table 4 are substantial (represented by the range between the MLE and Med2100 simulations). Indeed, the sizable uncertainties associated with human decision-making, climate mitigation, and adaptation are given as reasons for the EPA Final Rule to avoid estimating actual economic benefits or on-the-ground impacts from reduced greenhouse gas emissions. However, previous work has demonstrated, for example in the case of managing coastal risk in New Orleans, Louisiana, that in the face of climate change, the most expensive strategy we can pursue is to do nothing[27].

## 5 Code and Data Availability

All model and analysis code, input, and output files are available at https://zenodo.org/records/19577321. Figures and tables from this work may be reused or adapted with permission from the author.

## 6 Acknowledgments

I thank Stephane Sartzetakis from the Environmental Defense Fund for providing the FaIR temperature and ocean heat simulation output for the four scenarios examined here. This manuscript constitutes the academic work of the author. The views and opinions expressed in this work do not reflect those of Rochester Institute of Technology or the Environmental Defense Fund.## 7 References

1. EPA. *Rescission of the Greenhouse Gas Endangerment Finding and Motor Vehicle Greenhouse Gas Emission Standards Under the Clean Air Act*. 7686–7796 https://www.govinfo.gov/content/pkg/FR-2026-02-18/pdf/2026-03157.pdf (2026).
2. *Technical Memo on: Temperature, CO2 Concentration, and Sea Level Rise Impacts of Greenhouse Gas Emissions from U.S. Motor Vehicles for the "Rescission of the Greenhouse Gas Endangerment Finding and Motor Vehicle Greenhouse Gas Emission Standards Under the Clean Air Act" Final Rule*. https://www.regulations.gov/document/EPA-HQ-OAR-2025-0194-31105 (2026).
3. Wong, T. E. *et al.* BRICK0.2, a simple, accessible and transparent model framework for climate and sea-level projections. *Geosci. Model Dev.* **10**, (2017).
4. Levermann, A. *et al.* The multimillennial sea-level commitment of global warming. *Proc. Natl. Acad. Sci.* **110**, 13745–13750 (2013).
5. Clark, P. U. *et al.* Sea-level commitment as a gauge for climate policy. *Nat. Clim. Change* **8**, 653–655 (2018).
6. Nauels, A. *et al.* Multi-century global and regional sea-level rise commitments from cumulative greenhouse gas emissions in the coming decades. *Nat. Clim. Change* 1–7 (2025) doi:10.1038/s41558-025-02452-5.
7. Mengel, M. *et al.* Future sea level rise constrained by observations and long-term commitment. *Proc. Natl. Acad. Sci. U. S. Am. PNAS* **113**, 2597–2602 (2016).9